\documentclass{article}
\usepackage[preprint]{neurips_2026}

\usepackage[utf8]{inputenc} 
\usepackage[T1]{fontenc}    

\usepackage{hyperref}       
\usepackage{url}            

\usepackage{booktabs}       
\usepackage{amsfonts}       
\usepackage{nicefrac}       
\usepackage{microtype}      
\usepackage{xcolor}         

\usepackage{graphicx}
\usepackage{float}
\usepackage{bm}

\usepackage{amsmath} 
\usepackage{multirow}
\usepackage[ruled,vlined]{algorithm2e}
\SetKwInput{KwData}{Data}
\SetKwInput{KwInit}{Initialise}

\title{Cortico-cerebellar modularity as an architectural inductive bias for efficient temporal learning}

\author{
  Alexandra Voce \\
  Department of Bioengineering\\
  Imperial College London\\
  London, UK \\
  \texttt{a.voce25@imperial.ac.uk} \\
   \And
  Emmanouil Giannakakis \\
  Department of Bioengineering\\
  Imperial College London\\
  London, UK \\
  \texttt{e.giannakakis@imperial.ac.uk} \\
  \And
  Claudia Clopath \\
  Department of Bioengineering\\
  Imperial College London\\
  London, UK \\
  \texttt{c.clopath@imperial.ac.uk} \\
}

\begin{document}

\maketitle

\begin{abstract}
The cerebellum and cerebral cortex form tightly coupled circuits thought to support flexible and efficient temporal processing. How this interaction shapes cortical learning dynamics, and whether such heterogeneous modularity can benefit artificial systems, remains unclear. Here, we augment a recurrent neural network (RNN) with a cerebellar-inspired feedforward module and evaluate the resulting architecture on temporal tasks of varying difficulty. The cortico-cerebellar RNN (CB-RNN) learns faster and reaches higher maximum performance than parameter-matched fully recurrent baselines across a variety of regimes. Crucially, freezing the recurrent core after minimal training and delegating subsequent learning to the cerebellar module preserves superior learning efficiency, suggesting the cerebellar module is a primary driver of efficiency and that the cortical network can largely function as a fixed reservoir. Our results suggest that heterogeneous modular architectures can act as a powerful structural inductive bias in neural systems.
\end{abstract}

\section{Introduction}
Biological neural systems demonstrate greater learning efficiency, flexibility, and robustness than artificial systems~\cite{kudithipudi_biological_2022,lake_building_2017}. This disparity may stem from architectural organisation; biological intelligence arises from heterogeneous systems with distinct  architectures,  functions, and timescales, which can serve as an inductive bias to promote efficient and generalisable learning~\cite{goyal_inductive_2022,marblestone_toward_2016, salatiello_modularity_2026, kiebel2008hierarchy}.

The cortico-cerebellar system offers a well-motivated instance of such heterogeneous modularity. Cortical circuits support recurrent, context-dependent computation over time~\cite{miller_integrative_2001,mante_context-dependent_2013}, while the cerebellum (CB) is primarily a feedforward expansion-compression circuit, suited to rapid transformation and adaptive correction~\cite{marr_theory_1969,cayco-gajic_re-evaluating_2019,johansson_memory_2014}. These systems interact through closed reciprocal loops~\cite{middleton_cerebellar_2001}, and evidence from CB perturbation studies demonstrate that disrupting this interaction alters cortical activity and impairs cognitive performance~\cite{gao_cortico-cerebellar_2018,deverett_cerebellar_2019,verpeut_cerebellar_2023}. We propose that modifying a recurrent cortical network with a CB-inspired system could enhance learning efficiency and task performance by distributing learning across diverse modules.

Here, we test this hypothesis by augmenting a cortex-inspired Elman-type recurrent neural network (RNN)~\cite{elman_finding_1990} with a CB-inspired feedforward module that provides a learned additive bias signal to the recurrent core. Our results show that the CB-RNN learns faster, achieves higher task difficulty levels, and utilises parameters more efficiently than matched RNN-only baselines. These advantages persist even with reduced recurrent plasticity, indicating the CB module as the primary driver of task performance.  Representational analyses indicate distinct dimensionality and timescale profiles across RNN and CB modules, along with a task-dependent division of labour. Post-training ablations further confirm that the CB bias actively structures recurrent representations.

\section{Related Work}
\paragraph{Biological inductive biases, modularity, and heterogeneity.}
It is increasingly suggested that biological intelligence relies on architectural inductive biases that shape representations and learning dynamics within a system~\cite{marblestone_toward_2016,goyal_inductive_2022}. Modularity and heterogeneity can mitigate interference between learned functions, enhance robustness, and facilitate more efficient computation reuse~\cite{goyal_inductive_2022,salatiello_modularity_2026}. Evidence demonstrates that heterogeneity in neuronal properties boosts learning stability and robustness in spiking networks~\cite{perez-nieves_neural_2021}, while biologically inspired augmentations of recurrent networks can modify learning dynamics and representational geometry~\cite{soo_training_2023}. Learned modulatory signals, such as biases, can shape recurrent dynamics and facilitate flexible task performance without full synaptic plasticity~\cite{mason_versatile_2026,williams_expressivity_2024}. Collectively, these findings suggest that architectural organisation, rather than just parameter count or learning rule, is an important determinant of learning efficiency in neural systems.

\paragraph{Cortico-cerebellar circuits as heterogeneous learning systems.}
The cortico-CB system provides a concrete instance of heterogeneous modular interaction. Cortical circuits facilitate recurrent, context-dependent computation and long-term temporal integration~\cite{mante_context-dependent_2013,miller_integrative_2001,caligiore_consensus_2017}, while the CB is characterised by largely feedforward connectivity~\cite{tanaka_cerebro-cerebellum_2020,marr_theory_1969}. The CB input layer expands incoming mossy fibre signals through numerous granule cells (GCs), generating a high-dimensional sparse code conducive to pattern separation and rapid adaptive transformation over both sensory inputs and signals reflecting ongoing cortical activity~\cite{marr_theory_1969,cayco-gajic_re-evaluating_2019,tanaka_cerebro-cerebellum_2020,person_corollary_2019,rancz_high-fidelity_2007}, before being compressed through Purkinje cell (PC) outputs to deep cerebellar nuclei, completing closed cortico-CB loops~\cite{middleton_cerebellar_2001}. Experimentally, CB perturbation disrupts cortical dynamics and impairs working memory, sequence processing, and flexible behavioural updating~\cite{gao_cortico-cerebellar_2018,deverett_cerebellar_2019,verpeut_cerebellar_2023,stoodley_adaptive_2021}, indicating its active role in shaping cognition and behaviour through modulation of cortical computations.

\paragraph{Computational models of cortico-cerebellar interaction.} 
Recent modelling has begun to elucidate how CB-like modules can enhance learning in recurrent cortical networks. Boven et al.~\cite{boven_cerebro-cerebellar_2023} demonstrated that CB feedback can accelerate cortical learning by predicting feedback signals, thus alleviating delayed temporal credit assignment challenges. Pemberton et al.~\cite{pemberton2024cerebellar} revealed that CB-driven cortical dynamics facilitate task acquisition, switching, and consolidation, with CB signals remaining effective even when cortical weights are static, suggesting that the CB can drive cortical computation without destabilising learned representations. Agueci and Cayco-Gajic~\cite{agueci_distributed_2025} further indicated that distributing plasticity across a rapid CB system and a slower cortical network improves motor adaptation by enabling quick error corrections via the CB, while slower cortical plasticity ensures long-term retention of adapted behaviours. Together, these studies demonstrate that CB pathways can enhance cortical learning when given explicit predictive or error-corrective targets. 

\section{Methods}
\begin{figure}[htbp]
\centering
\includegraphics[width=0.95\textwidth]{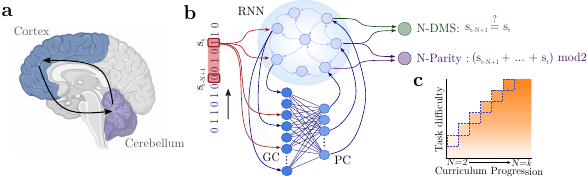}
\vspace{-0.5em}
    \caption{\textbf{The cortico-cerebellar RNN (CB-RNN) architecture and curriculum design.}
    (a) Cortico-CB loop motivating the model architecture. (b) The CB-RNN combines a recurrent core with a CB-inspired feedforward module, which receives the current input and recurrent hidden state, processes them through GC- and PC-like layers, and returns a learned bias to the RNN. Models trained on N-DMS and N-Parity tasks. (c) Curriculum learning progressively increases task difficulty from $N=2$ to $N=k$, with outlined boxes denoting the current curriculum level.}
 \label{Model_fig}
\end{figure}
\subsection{Network Model}
\paragraph{Baseline RNN.}
Compute resources are provided in Appendix~\ref{app:compute}. The cortical component is an Elman-type recurrent neural network~\cite{elman_finding_1990}. Given input $x_t \in \mathbb{R}^{d}$ and hidden state $h_t \in \mathbb{R}^{N_h}$, the model updated its activity as
\begin{align}
h_{t+1} &= (1-\alpha)h_t + \alpha\,\phi(W_hh_t + W_xx_t + b), 
\end{align}
where $W_h$ and $W_x$ are recurrent and input weights, respectively, and $b$ is a recurrent bias. The network's readout is given by $y_t = W_{hy}h_t$, where $W_{hy}$ is a trainable linear readout. The leak parameter $\alpha = 1/\tau$ controls the effective integration timescale and was fixed at $\tau = 1.5$. The activation function was a Leaky ReLU, denoted $[\cdot]_{\gamma}$ with a negative slope ($\gamma=0.01$)~\cite{maas_rectifier_2013}.
RNN-only baselines were scaled to match the number of trainable parameters in each corresponding CB-RNN variant, allowing comparisons at approximately matched model capacity (see Appendix~\ref{app:parameter-matching}). GRU-based controls are reported in Appendix~\ref{app:GRU}.
\paragraph{Cortico-cerebellar model.}
The CB-RNN augments the baseline recurrent network with an additional feedforward module designed to provide a CB-inspired modulatory signal to the recurrent core (Fig.~\ref{Model_fig}). In all main experiments, the recurrent hidden size was fixed at 64 units. The CB module used a GC expansion layer of 256 units; performance and representational analyses for additional GC sizes (64, 128, 512) are presented in Appendix~\ref{app:other_sizes}.

At each timestep, the CB module received a concatenated input $c_t = [h_t;x_t]$, comprising both the recurrent hidden state $h_t$ and the task input $x_t$, motivated by evidence that the CB receives convergent information from the cortex and sensory pathways~\cite{rancz_high-fidelity_2007,person_corollary_2019}. Control comparisons using $h_t$ and $x_t$ alone are reported in Appendix~\ref{app:input_controls}. The CB module consisted of two nonlinear feedforward transformations followed by a linear output projection:
\begin{align}
z_t &= \phi_{\mathrm{CB}}(W_{\mathrm{GC}}c_t + b_{\mathrm{GC}}), \\
u_t &= \phi_{\mathrm{CB}}(W_{\mathrm{PC}}z_t + b_{\mathrm{PC}}), \\
b_{\mathrm{cb},t} &= W_{\mathrm{out}}u_t + b_{\mathrm{out}} .
\end{align}
Here, $z_t$ and $u_t$ denote the activity of the expanded GC-like and PC-like stages, respectively. The final linear projection produces the CB bias ($b_{\mathrm{cb},t}$), which is returned to the recurrent core. ReLU nonlinearities were applied in the GC-like and PC-like stages. The recurrent update in the CB-RNN was therefore:
\begin{equation}
h_{t+1} = (1-\alpha)h_t + \alpha [W_hh_t + W_xx_t + b + b_{\mathrm{cb},t}]_{\gamma},
\end{equation}
where $b_{\mathrm{cb},t}$ denotes the CB contribution and $[\cdot]_{\gamma}$ denotes the Leaky ReLU nonlinearity. Thus, the CB modulates cortical dynamics by adding a learned, task-dependent bias to the recurrent update. This makes the recurrent update conditional on a learned feedforward transformation, relating the CB-RNN to recent architectures that augment recurrent networks with feedforward components to enrich recurrent hidden-state dynamics and improve sequence learning~\cite{pascanu_how_2014,sun2024learning}.

\subsection{Task Suite}
Models were evaluated on two binary sequence-processing tasks commonly used to benchmark recurrent memory~\cite{khajehabdollahi_emergent_2023}: $N$-bit delayed match-to-sample (DMS) and $N$-bit Parity (Parity). Across tasks, models received variable-length binary sequences, with sequence length sampled as a function of the largest active $N$. 

\paragraph{N-Bit Delayed Match-to-Sample.}
In DMS, the model received a binary input sequence and was required to report whether the final input bit matched the bit presented $N-1$ timesteps earlier. The task requires the model to retain inputs in memory, thus, increasing $N$ increases memory demands.

\paragraph{N-Bit Parity.}
In the Parity task, the model was required to report the parity of the final $N$ input bits, i.e. whether the number of ones in the previous $N$-bit window was odd or even. Unlike DMS, which depends on comparison with a specific past input, parity requires nonlinear integration over the full temporal window.

\subsection{Training}
\subsubsection{Optimisation}
Models were trained using stochastic gradient descent (SGD) with Nesterov momentum ($\mu = 0.1$) and a cross-entropy loss objective. Gradients were clipped to a maximum $\ell_2$ norm of $7.5$. Training epochs consisted of 100 gradient steps over batches of 64 sequences, with performance evaluated over 50 held-out batches at the end of each epoch. Learning rates were set for the RNN and CB modules at $\eta = 0.01$. RNN weights were initialised using Kaiming-uniform. The hidden state was initialised as $h_0 \sim \mathcal{U}(0, 0.1)$ per batch. For CB-RNN models, CB biases were initialised to zero; input$\to$GC weights were sampled from $\mathcal{N}(0, 0.01^2)$, and GC$\to$PC and PC$\to$DCN weights from $\mathcal{N}(0, 0.1^2)$. Reported results were repeated across random seeds ($n=3$). Full task-generation and training details are provided in Appendix~\ref{app:task_generation}.

\subsubsection{Curriculum Learning}
Task difficulty was controlled by the delay/window parameter $N$. Training commenced at $N=2$ and advanced incrementally once performance exceeded 98\% accuracy at test~\cite{bengio_curriculum_2009, khajehabdollahi_emergent_2023}. Three variants for CB-RNN models were evaluated. In the \textit{simultaneous} variant, all modules trained concurrently. In the \textit{full reservoir} variant, the recurrent core was frozen after  training on $N=2$, with all subsequent learning carried by the CB module and readout heads. In the \textit{interleaved reservoir} variant, recurrent plasticity was additionally reinstated at fixed curriculum intervals: the RNN was trained whenever $(N-2) \bmod k = 0$ and refrozen once that level was solved, with $k=10$ for DMS and $k=5$ for Parity to account for their different curriculum ranges.

\subsubsection{Training Regimes}
\paragraph{Single-Task Training.} In single-task training, each model was trained on one task at a time, with $N$ increasing by one upon surpassing 98\% evaluation accuracy. Training proceeded until $N=100$ or the 1000-epoch limit was met. Performance was assessed using three metrics: the maximum $N$ achieved within the training budget; the area under the curriculum progression curve (AUC), indicating cumulative learning progress; and the epochs needed to reach half of the global maximum $N$ attained across all models ($N_{\mathrm{global}}/2$), reflecting learning speed.

\paragraph{Multi-Task Training}
In multi-task training, models solved DMS and Parity simultaneously from shared input sequences, with separate task-specific readout heads each trained on a cross-entropy loss. Curriculum progression was synchronised across tasks, advancing only when both tasks exceeded the 98\% threshold at the current $N$. The same metrics were used to summarise performance as in single-task training.

\paragraph{Task-Switch Training}
Task-switching experiments alternated training between the DMS and Parity tasks across successive phases. In each phase, only the active task's readout head was trained, while both tasks were evaluated after every epoch. Task switches occurred when the active task reached the next multiple of $N_{\mathrm{switch}} =5$, with the incoming task resuming directly at that $N$ rather than restarting from $N=2$. Consequently, this approach introduced the incoming task at a novel difficulty level, requiring adaptation without complete reacquisition. Curriculum progression ceased after approximately 600 epochs, and all analyses related to task switching were performed over the initial 650 epochs. AUC values were derived from task-specific curriculum trajectories and averaged across tasks. The speed of post-switch adaptation was measured as the mean number of epochs needed to progress from the phase-start $N$ to $N+1$ after a switch, averaged across DMS and Parity phases.

\subsection{Activity Analysis}
\paragraph{Population Dimensionality Analysis}
\label{section:ActivityPCA}
Representational dimensionality was assessed using principal component analysis (PCA) on neural activity from saved model checkpoints, which included learned weights and biases after each curriculum level (see Appendix~\ref{app:pca_details}). For each run and curriculum level $N$, the relevant checkpoint was evaluated on 10 held-out batches of 64 sequences. Activity was extracted separately for each population of interest: the recurrent hidden state $h_t$, GC layer $z_t$, PC layer $u_t$, and CB bias signal $b_{cb,t}$. The collected activities were flattened across time and batch dimensions, mean-centred, and PCA was applied to each population's resulting matrix. Dimensionality was summarised by the number of principal components needed to explain 95\% of the variance ($d_{95}$). This process was repeated across all curriculum levels, random seeds, tasks, curriculum variants, and populations. 

\paragraph{Intrinsic Timescale Analysis}
Intrinsic timescales were estimated from population autocorrelations under stationary random input. At each $N$ checkpoint, the model was driven by $T=20{,}000$ timesteps of i.i.d.\ Bernoulli(0.5) input across $B=8$ independent streams, discarding the first 500 timesteps as burn-in. A population activity trace was formed per stream by summing across units, and autocorrelations were estimated using a Fast Fourier Transform (FFT)-based estimator, corrected for lag-dependent sample overlap, averaged across streams, and normalised by the zero-lag value. Population timescale $\tau_{pop}$ was defined as the first lag at which the normalised autocorrelation fell below $e^{-1}$, evaluated up to a maximum lag of 200 timesteps~\cite{zeraati_flexible_2022}.

\paragraph{Cerebellar Ablation Analysis}
\label{methods:ablation}
To test whether trained CB-RNN models relied on the CB pathway for task performance, the CB bias $b_{cb}$ was silenced post-training without retraining any weights, and accuracy was re-evaluated across all curriculum levels $N$ over 20 held-out batches of 64 sequences. The effect of ablation was quantified as $\Delta\mathrm{Acc} = \mathrm{Acc}_{\mathrm{full}} - \mathrm{Acc}_{\mathrm{ablated}}$, with positive values indicating performance supported by the CB pathway. T-distributed stochastic neighbour embedding (t-SNE) was implemented for visualisation of output class representations in the recurrent core (see Appendix~\ref{app:tsne_details} for details). For the visualised $N$, class separability was quantified using a linear decoder (Appendix~\ref{app:class_separability}). To verify that accuracy drops were not attributable to a mismatch between recurrent dynamics and readout weights trained in the presence of $b_{cb}$, a readout-retraining control was also performed (see Appendix~\ref{app:readout_retraining}). 

\paragraph{Code Availability} The code for all simulations is available in 
\href{https://anonymous.4open.science/r/Cortico-cerebellar_RNNs/}{anonymous.4open.science/r/Cortico-cerebellar\_RNNs}.

\section{Results}
\subsection{CB-RNNs outperform matched RNN-only baselines in efficiency and final performance}
We first tested whether CB-inspired modularity improved learning efficiency in single-task curricula. Models were trained on tasks whose difficulty was indexed by the temporal span $N$, with $N$ increasing when performance exceeded the advancement criterion. CB-RNNs were compared with RNN-only baselines that had equivalent trainable parameter counts to ensure that learning capacity was consistent across the comparisons.

\begin{figure}[htbp]
    \centering
    \includegraphics[width=1\textwidth]{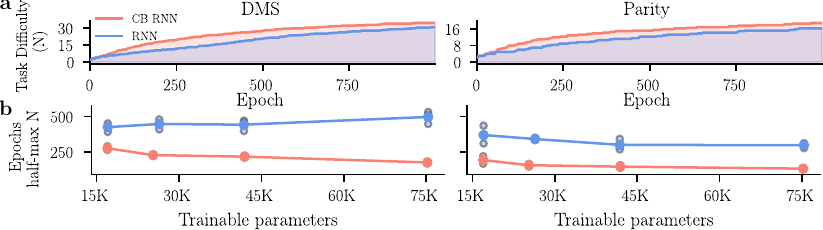}
    \vspace{-1.6em}
    \caption{\textbf{CB-RNN modularity improves curriculum learning efficiency and final performance.} (a) Task difficulty $N$ reached over training for DMS (left) and Parity (right); shading denotes AUC. (b) Epochs to half of the global maximum $N$ across model scales. CB-RNN models consistently learned faster and reached higher difficulty levels than parameter-matched RNN-only baselines at all model sizes.}
    \label{fig:fig2}
\end{figure}

Across both tasks, CB-RNNs progressed through the curriculum more efficiently than parameter-matched RNN-only baselines (Fig.~\ref{fig:fig2}a), reaching higher final $N$ values, accumulating larger AUCs, and approaching half of the global maximum $N$ roughly twice as fast (see Appendix~\ref{app:performance-statistics}). This advantage was also consistent across model scales: CB-RNNs reached half-maximal task difficulty earlier than matched RNN-only baselines at every model size examined, and smaller CB-RNNs matched or exceeded the performance of larger RNN-only models (Fig.~\ref{fig:fig2}b). Together, these results demonstrate that CB-inspired modularity improves both the speed and parameter efficiency with which recurrent networks acquire sequential tasks.

\subsection{CB-RNN advantage persists under multi-task and task-switching demands}
\begin{figure}[htbp]
    \centering
    \includegraphics[width=1\textwidth]{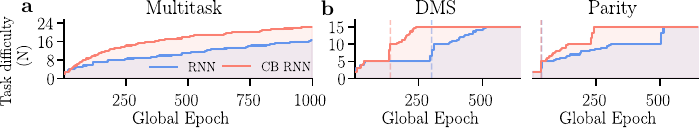}
    \vspace{-1.4em}
    \caption{\textbf{CB-RNN models retain learning advantage under multi-task and task-switching demands.} (a) Curriculum progression during multi-task training; shading denotes AUC. (b) Phase-resolved curriculum trajectories during task switching; flat segments reflect epochs where the other task was active, dashed lines mark switch epochs. CB-RNNs accumulated more learning progress and adapted faster following each switch than RNN-only baselines.}
    \label{fig:finding2}
\end{figure}
\paragraph{Multi-task.}
We next asked whether the CB-RNN advantage persisted when models were required to solve both tasks simultaneously. Under multi-task curriculum training, CB-RNNs again progressed faster, reached higher final $N$ values, and accumulated larger AUCs than parameter-matched RNN-only baselines (Fig.~\ref{fig:finding2}a; Appendix~\ref{app:performance-statistics}). Notably, the absolute magnitude of the CB-RNN advantage was more pronounced in the multi-task setting than in single-task training, indicating that the architecture may be particularly advantageous when the recurrent core is required to concurrently fulfil multiple task demands. 

\paragraph{Task switching.}
To assess whether CB-RNNs could more flexibly adapt to a new task objective at an inherited difficulty level, models were trained in alternating phases on the tasks, with each switch presenting the incoming task at a previously unseen difficulty level. CB-RNNs accumulated greater cumulative learning progress than RNN-only baselines across the full switching experiment (AUC: $5{,}644 \pm 13$ vs.\ $3{,}187 \pm 83$; Fig.~\ref{fig:finding2}b), and adapted roughly twice as fast following each switch ($31.00 \pm 6.06$ vs.\ $59.67 \pm 10.60$ epochs to advance one difficulty level from phase-start $N$). This suggests that, in addition to improving learning progression speed, the CB module supports more efficient transfer of learned sequence representations across task objectives.

\subsection{CB-RNN efficiency benefit persists under restricted recurrent plasticity}
\begin{figure}[htbp]
    \centering
    \includegraphics[width=1\textwidth]{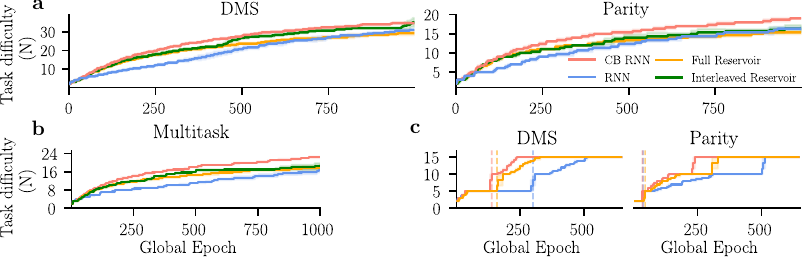}
    \vspace{-1.7em}
    \caption{\textbf{Cerebellar module drives learning efficiency even when RNN plasticity is restricted.} (a) Single-task and (b) multi-task curriculum progression; (c) task-switching trajectories. Reservoir variants outpaced the RNN-only baseline, demonstrating that the CB module is a primary driver of the efficiency advantage.}
    \label{fig:finding3}
\end{figure}

To investigate whether the CB-RNN efficiency advantage depended on continuous optimisation of both the recurrent core and CB module, we trained two reservoir variants that restricted recurrent plasticity: a full reservoir model, in which the RNN was frozen after initial training, and an interleaved reservoir model, in which recurrent learning was periodically reinstated at fixed intervals (Fig. ~\ref{fig:finding3}a). 
Under single-task and multi-task curricula, both reservoir variants outperformed the RNN-only baseline in AUC and speed of approach to the global half-maximum $N$, despite having a largely frozen recurrent core (Fig.~\ref{fig:finding3}b--c; Appendix~\ref{app:performance-statistics}, Table~\ref{tab:performance_stats_table}). The full reservoir fell slightly below the RNN-only baseline in final $N$ reached, consistent with the cost of extended recurrent freezing at high difficulty levels, while the interleaved reservoir model recovered this deficit on both tasks. The fully trained CB-RNN remained the strongest model across metrics.
In the task-switching paradigm, the full reservoir again performed above the RNN-only baseline but below the CB-RNN across all metrics (AUC: $4{,}984 \pm 134$ vs.\ $3{,}187 \pm 83$ and $5{,}644 \pm 13$; post-switch adaptation: $40.33 \pm 2.31$ vs.\ $59.67 \pm 10.60$ and $31.00 \pm 6.06$ epochs; Fig.~\ref{fig:finding3}d), indicating that recurrent plasticity contributes, but is not required for CB efficiency benefit to emerge. 

\subsection{Analysis of Network Dynamics}
\begin{figure}[htbp]
    \centering
    \includegraphics[width=1\textwidth]{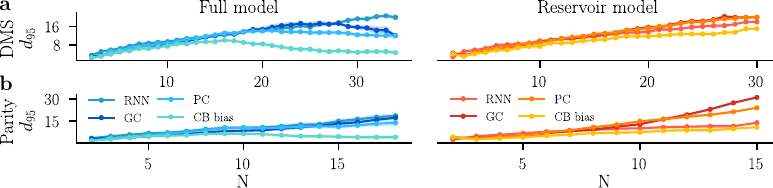}
    \vspace{-1.7em}

    \caption{\textbf{Restricted recurrent plasticity shifts representational expansion across model modules.} $d_{95}$ across task difficulty $N$ for DMS (a) and Parity (b), in full (left) and reservoir (right) models. In the full model, CB populations compress at higher $N$s while RNN hidden state continues to expand; this compression is absent in reservoir models, where CB modules instead sustain representational growth. Lines show means; shading indicates SD.}
    \label{fig:finding4}
\end{figure}

\paragraph{Restricted recurrent plasticity shifts representational expansion onto the cerebellar pathway}
To characterise how each module's representational geometry evolved with task difficulty, we tracked $d_{95}$ across
$N$ for each module in the full CB-RNN and full reservoir models (Fig.~\ref{fig:finding4}).

In the full CB-RNN, all modules initially expanded their representational dimensionality with increasing $N$ on both tasks. In DMS, the modules diverged at higher $N$: the CB bias output began decreasing first, becoming the lowest-dimensionality population at large $N$s, with compression subsequently emerging in PC and GC layers at higher $N$s (Fig.~\ref{fig:finding4}a). The recurrent hidden state continued to expand throughout. For Parity, only the CB bias compressed at higher $N$s, while all other populations continued to increase (Fig.~\ref{fig:finding4}b). In the reservoir, this compression was absent across both tasks, with all modules expanding as $N$ increased, and CB populations GC and PC reaching higher dimensionality than the frozen recurrent hidden state.

These results suggest a task-dependent division of labour between RNN and CB modules. For DMS, the easier of the two tasks, compression propagated throughout all CB populations at higher $N$s while the recurrent hidden state sustained representational expansion. For the Parity task, only the CB bias compressed, suggesting that greater computational demands may require more sustained CB support. In both cases, removing recurrent plasticity abolished this compression, with CB populations expanding to account for the representational burden that the frozen RNN cannot provide.

\begin{figure}[htbp]
    \centering
    \includegraphics[width=1\textwidth]{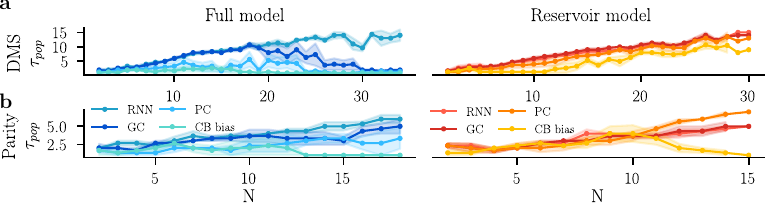}
    \vspace{-1.5em}
    
    \caption{\textbf{Reservoir constraints redistribute intrinsic timescales across recurrent and cerebellar modules.} Population timescale ($\tau_{pop}$) across task difficulty $N$ for DMS (a) and Parity (b), in full (left) and reservoir (right) models. In the full model, the RNN hidden state dominates long timescales at high $N$s, while CB populations decline; in reservoir models, timescales redistribute across CB modules to compensate for the frozen recurrent core. Lines show means; shading indicates SD.}
    \label{fig:finding5}
\end{figure}

\paragraph{Reservoir constraints redistribute longer intrinsic timescales across modules}
We subsequently examined the evolution of intrinsic timescales across modules as task difficulty increased (Fig.~\ref{fig:finding5}). The population timescale $\tau_{\mathrm{pop}}$ represents the lag at which population autocorrelation decays to $1/e$, and was monitored across varying task difficulties $N$.

In the full CB-RNN on DMS, the recurrent hidden state and CB GC layer developed the longest timescales, increasing together until $N \approx 18$, after which the CB populations gradually declined, leaving the recurrent core as the dominant long-timescale module at high $N$s (Fig.~\ref{fig:finding5}a). The CB bias remained lowest throughout. For Parity, the pattern was similar but weaker, with the RNN showing the clearest increase and the CB bias declining at the higher $N$s (Fig.~\ref{fig:finding5}b).

For the reservoir models, this division did not emerge. For DMS, all populations increased with $N$, and though the CB bias remained the lowest, it was no longer flat. For Parity, the PC layer became the highest-timescale population, overtaking the RNN and GC layer, while the CB bias again declined at higher $N$s. Together, these timescale findings closely mirror the prior dimensionality results: in the full model, the recurrent core is the primary substrate for long timescales at high $N$s, while removing its plasticity redistributes this temporal structure across the CB. 

Together, these findings suggest that reservoir constraints shift the burden of temporal representation onto the CB pathway. Whereas full CB-RNNs can rely on an adaptable recurrent core to sustain long-timescale dynamics, reservoir models appear to compensate by recruiting longer timescales across CB transformations.

\subsection{Cerebellar ablation disrupts task performance and recurrent representations}

\begin{figure}[htbp]
    \centering

    \begin{minipage}[c]{0.14\textwidth}
        \centering
        \raisebox{0.28\height}{%
            \includegraphics[width=\textwidth]{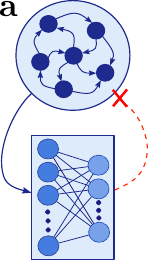}
        }
    \end{minipage}
    \hfill
    \begin{minipage}[c]{0.84\textwidth}
        \centering
        \includegraphics[width=\textwidth]{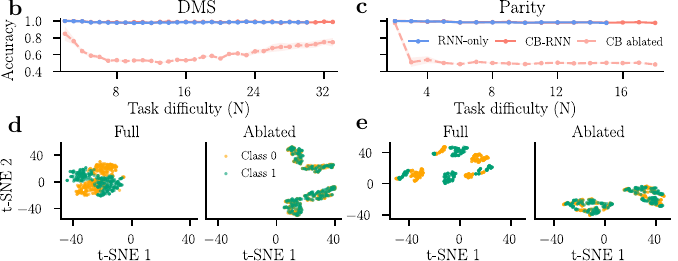}
    \end{minipage}

    \caption{\textbf{Cerebellar bias is necessary for task performance and supports class-discriminative representations.}
    (a) Ablation schematic. (b, c) Accuracy across $N$ for full, CB-ablated, and RNN-only models on DMS and Parity. CB ablation substantially reduced accuracy, with partial recovery on DMS but not Parity. Lines show mean across runs; shading indicates SD. (d, e) t-SNE projections of recurrent hidden-state activity by output class at representative $N$ values, with and without CB bias. Class separation is reduced after ablation.}
    \label{fig:finding6}
\end{figure}
To test whether trained CB-RNN models relied on the CB pathway, we silenced $b_{cb}$ post-training without retraining any weights, and re-evaluated accuracy across all curriculum levels $N$ (Fig.~\ref{fig:finding6}a--c). On DMS, ablation produced a substantial accuracy drop across all $N$s (mean $\Delta\mathrm{Acc} = 0.37 \pm 0.09$) (Fig.~\ref{fig:finding6}b), with partial recovery at higher $N$s but no return to full model performance. This partial recovery is consistent with the prior dimensionality and timescale analyses, which demonstrate that CB contributions compress at higher $N$s as the recurrent core increasingly dominates task performance. For Parity, the effect of ablation was more severe, reducing accuracy to near chance across all $N$s (mean $\Delta\mathrm{Acc} = 0.45 \pm 0.11$), with no recovery, indicating that the recurrent core alone cannot support the complex temporal integration required by this task once trained with CB bias (Fig.~\ref{fig:finding6}c). To confirm that these effects were not attributable to a readout--recurrent mismatch, readout heads were permitted to retrain following ablation; accuracy did not recover, confirming a genuine dependence on the CB (see Appendix~\ref{app:readout_retraining}).

Finally, to visualize the representational consequences of ablation, we applied t-SNE on recurrent activity at representative difficulty levels (DMS: $N=12$; Parity: $N=6$). In the full model, classes formed separated clusters on both tasks; ablation caused this separation to collapse, with classes becoming intermixed in the recurrent core (Fig.~\ref{fig:finding6}d--e, Appendix~\ref{app:class_separability}). This indicates that the CB bias actively structures recurrent representations to be class-discriminative. 

\section{Discussion} 
\paragraph{Cortico-cerebellar modularity as an inductive bias for learning.} 
Our findings show that cortico-CB-inspired modularity improves learning efficiency through architectural heterogeneity rather than parameter count. CB-RNNs consistently outperformed parameter-matched baselines in both efficiency and final performance, supporting the role of modular inductive biases in constraining learning and accelerating convergence~\cite{marblestone_toward_2016,salatiello_modularity_2026,goyal_inductive_2022}. Reservoir experiments corroborate this, finding that CB bias learning with a fixed recurrent core learns faster than fully plastic RNNs, highlighting the CB pathway’s contribution. The added benefit of periodic recurrent plasticity suggests a functional division in which the CB supports rapid adaptation while the cortex supports slower temporal integration, consistent with fast–slow cortico-CB models~\cite{caligiore_consensus_2017, tanaka_cerebro-cerebellum_2020, agueci_distributed_2025} and prior work on learned biases in ANNs~\cite{mason_versatile_2026,williams_expressivity_2024}. The CB-RNN advantage was strongest in multi-task settings, where competing representational demands arise, aligning with evidence linking dual-task performance to cortico-CB connectivity~\cite{muller_parallel_2023}. CB-RNNs also adapted about twice as fast after task switches, suggesting more transferable representations, consistent with the CB’s role in flexible behavioural adaptation~\cite{verpeut_cerebellar_2023, peterburs_role_2018,stoodley_altered_2017}. Overall, these results indicate that modularity enhances learning and generalisation in complex, multi-objective settings~\cite{marblestone_toward_2016, salatiello_modularity_2026,goyal_inductive_2022}.

\paragraph{Emergent division of labour and functional necessity of the cerebellar pathway.} 
Modular specialization emerged through learning rather than design constraints. CB populations operated on faster timescales than the recurrent core, suggesting that structural differences between feedforward CB and recurrent cortical circuits naturally induce timescale separation, consistent with modular efficiency~\cite{salatiello_modularity_2026,goyal_inductive_2022} and known cortico-CB hierarchies~\cite{caligiore_consensus_2017, tanaka_cerebro-cerebellum_2020, agueci_distributed_2025}. Dimensionality analyses showed that CB representations expanded then compressed with task difficulty (in DMS), while the recurrent core continued to expand; this redistribution, absent in reservoir models, indicates active modular specialisation~\cite{marblestone_toward_2016, goyal_inductive_2022}. Ablation confirmed functional necessity: silencing the CB pathway impaired performance and reduced class separability in recurrent representations, implying cortical computations partly depend on CB input, in line with experimental evidence~\cite{israely_cerebellar_2025, gao_cortico-cerebellar_2018}. Finally, task-dependent differences, where CB compression and partial recovery after ablation occurred only for the DMS task, support load-dependent recruitment~\cite{kirschen_load-_2005} and a consolidation view in which the CB supports acquisition while the cortex gradually internalises solutions~\cite{herzfeld_contributions_2014, pemberton2024cerebellar, agueci_distributed_2025}.

\paragraph{Limitations.}
Various modelling choices influence the interpretation and scope of these findings. The model is a coarse abstraction that omits biological details such as CB-specific teaching signals and subcortical relays via the pons and thalamus~\cite{ros_population_2025, hoang_dynamic_2023, habas_cerebellar_2019, ito_climbing_1982}. The tasks and recurrent core are deliberately simple, GRU controls showed an attenuated CB benefit (Appendix~\ref{app:GRU}), and the use of backpropagation through time lacks biological plausibility~\cite{lillicrap_backpropagation_2020}, limiting generalisation to richer settings. Future work should test whether the observed efficiency gains persist with greater biological realism, including CB-specific plasticity, more complex models, and more plausible learning rules~\cite{murray_local_2019,bellec_solution_2020,agueci_distributed_2025}.

\paragraph{Future directions.}
These findings suggest several directions for future work. The model predicts that CB dynamics should evolve faster than frontal cortical dynamics during working memory tasks, especially under higher cognitive load, and that CB perturbation should disrupt the class-separable geometry of cortical representations. CB populations are also expected to exhibit load-dependent dimensionality expansion during acquisition followed by compression at later stages. Future studies can test whether these efficiency gains generalise to more complex architectures and reinforcement learning agents, and use combined top-down and bottom-up approaches to determine which biological details are essential for preserving the cortico-CB advantage.

\bibliographystyle{unsrt}
\bibliography{bibliography}

\newpage
\appendix
\section{Compute resources.}
\label{app:compute}
All experiments were run on a local workstation equipped with three NVIDIA GPUs: one GeForce GTX TITAN X and two GeForce RTX 2080 Ti GPUs. Experiments used CUDA 12.6. All simulations and analyses were run with Python 3.9.6, PyTorch 2.8.0, NumPy 2.0.2, scikit-learn 1.6.1, and Matplotlib 3.9.4. The code is available in 
\href{https://anonymous.4open.science/r/Cortico-cerebellar_RNNs/}{anonymous.4open.science/r/Cortico-cerebellar\_RNNs}.

\section{Parameter matching between recurrent-only and CB-RNN models}
\label{app:parameter-matching}

To ensure that performance differences between recurrent-only and CB-RNN models could not be attributed to different trainable parameter counts, recurrent-only baselines were matched to CB-RNN models by total trainable parameter count. Matching was performed separately for Elman and GRU architectures, as their recurrent cores differ in parameter count.

\paragraph{Elman RNN.}
For an Elman RNN with input size $I$ and hidden size $H$, the recurrent core contains input-to-hidden and hidden-to-hidden weights and biases:
\[
P_{\mathrm{core}}^{\mathrm{Elman}} = IH + H + H^2 + H.
\]

\paragraph{GRU.}
For a GRU with input size $I$ and hidden size $H$, the recurrent core contains update, reset, and candidate pathways, each with input weights, recurrent weights, and two bias vectors:
\[
P_{\mathrm{core}}^{\mathrm{GRU}} = 3\bigl(IH + H^2 + 2H\bigr).
\]

\paragraph{CB module and readout.}
For the reported CB-RNN models, the CB module received the concatenated input $c_t=[h_t;x_t]$ of dimension $H+I$. With GC dimension $G$ and PC and CB output dimensions fixed to the hidden size $H$, the CB module contributed
\[
P_{\mathrm{CB}} = (H+I)G + G + GH + H + H^2 + H.
\]
This corresponds to the GC transformation, the PC transformation, and the final CB output transformation back to hidden-state dimensionality.

The multi-head readout, with $M$ heads and $K$ classes, contributed
\[
P_{\mathrm{readout}} = M(HK + K),
\]
including readout biases. Total CB-RNN parameters were therefore
\[
P_{\mathrm{CB\text{-}RNN}} =
P_{\mathrm{core}} + P_{\mathrm{CB}} + P_{\mathrm{readout}},
\]
where $P_{\mathrm{core}}$ is the architecture-specific recurrent core term above.

For each CB-RNN configuration, the matched recurrent-only hidden size was selected as
\[
H^\ast =
\arg\min_H
\left|
P_{\mathrm{recurrent\text{-}only}}(H) -
P_{\mathrm{CB\text{-}RNN}}
\right|,
\]
where
\[
P_{\mathrm{recurrent\text{-}only}}(H)
=
P_{\mathrm{core}}(H) + M(HK+K).
\]
This produced recurrent-only baselines with approximately matched trainable capacity, so comparisons reflected architectural organisation rather than parameter count alone.

\section{GRU-based Controls}
\label{app:GRU}

To test whether the CB-RNN advantage depended on the use of an Elman recurrent core, the single-task DMS and Parity experiments were repeated using a gated recurrent unit (GRU) architecture. The GRU models were implemented to match the Elman experiments as closely as possible, including the same curriculum-learning procedure, task generation, optimisation settings, multi-head readout structure, CB module design, and parameter-matched recurrent-only comparisons.

The baseline GRU used standard update and reset gates, with a candidate hidden state computed from the current input and reset-gated previous hidden state. In the CB-GRU model, the cerebellar module received the same input configuration as in the main CB-RNN experiments and produced a hidden-sized bias vector. This bias was injected only into the candidate hidden-state preactivation, rather than into the update or reset gates:
\[
z_t = \sigma(W_z x_t + U_z h_{t-1}),
\]
\[
r_t = \sigma(W_r x_t + U_r h_{t-1}),
\]
\[
\tilde{h}_t = \phi(W_n x_t + U_n(r_t \odot h_{t-1}) + b_{\mathrm{cb},t}),
\]
\[
h_t = (1 - z_t) \odot h_{t-1} + z_t \odot \tilde{h}_t .
\]
Here, $\phi$ is the candidate activation function and $b_{\mathrm{cb},t}$ is the cerebellar bias. The CB bias was restricted to the candidate pathway to preserve the GRU gating mechanism while providing the closest architecture-specific analogue of the additive recurrent preactivation bias used in the Elman CB-RNN. The GRU controls also used an additional norm cap on the CB bias. This was included because the bias was injected into the candidate-state pathway of a gated recurrent transition, where large additive signals could saturate the candidate activation and interact strongly with the update gate. The cap therefore kept the CB contribution on the same scale as the GRU candidate preactivation.

These controls tested whether CB-inspired biasing improved temporal learning beyond the specific dynamics of an Elman RNN. Performance was compared between CB-GRU models and parameter-matched GRU-only baselines under the same single-task curriculum procedure used for the Elman models.

\begin{figure}[htbp]
    \centering
    \includegraphics[width=1\textwidth]{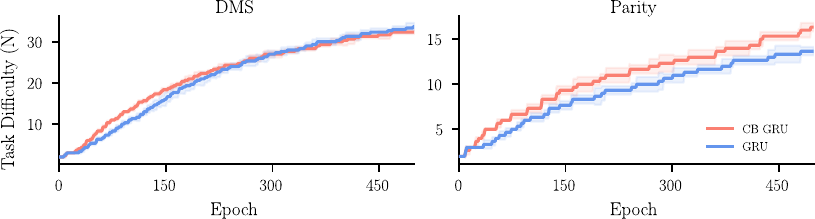}
    \vspace{-1.5em}
    \caption{\textbf{GRU-based controls reveal an attenuated and task-dependent CB benefit.}
    Single-task DMS and Parity curriculum progression was compared between CB-GRU models and parameter-matched GRU-only baselines. On DMS, CB-GRU showed faster early curriculum progression but reached a similar final difficulty to the GRU-only baseline. On Parity, CB-GRU progressed faster and reached higher final task difficulty. Lines show means across random seed repeats; shaded regions show SD.}
    \label{fig:GRU_results}
\end{figure}

Together, the GRU controls revealed an important boundary condition on the CB-RNN advantage (Fig.~\ref{fig:GRU_results}, Table~\ref{tab:gru_performance}). 

\begin{table}[h]
\centering
\caption{\textbf{Curriculum-learning performance for GRU-based single-task controls.} Values are mean $\pm$ SD across runs. Half-maximum epochs were measured relative to the global maximum $N$ across models within each task.}
\label{tab:gru_performance}
\begin{tabular}{llccc}
\toprule
Task & Model & Max $N$ & Epochs to $N_{\mathrm{global}}/2$ & AUC \\
\midrule
\multirow{2}{*}{\shortstack[l]{DMS}}
  & CB-GRU               & $32.33 \pm 0.58$ & $141.67 \pm 6.11$   & $11{,}009 \pm 264$ \\
  & GRU-only             & $33.67 \pm 1.15$ & $163.00 \pm 15.10$  & $10{,}700 \pm 480$ \\
\midrule
\multirow{2}{*}{\shortstack[l]{Parity}}
  & CB-GRU               & $16.33 \pm 0.58$ & $163.00 \pm 14.00$   & $5{,}389 \pm 227$ \\
  & GRU-only             & $13.67 \pm 0.58$ & $189.33 \pm 26.39$  & $4{,}602 \pm 267$ \\
\bottomrule
\end{tabular}
\end{table}
On DMS, the CB-GRU reached the global half-maximum difficulty faster and showed a slightly higher AUC than the GRU-only baseline, but reached a slightly lower final difficulty, indicating that the strong CB benefit observed with Elman RNNs was substantially attenuated in a gated recurrent architecture. On Parity, however, CB-GRU outperformed the GRU-only baseline across all summary metrics, reaching higher final $N$, progressing faster, and accumulating larger AUC. These results suggest that CB-inspired biasing can still improve temporal learning in GRUs, particularly under stronger nonlinear integration demands, but that the effect is not architecture-independent. Rather, the CB pathway may be most beneficial when the recurrent core lacks intrinsic gating mechanisms for regulating temporal integration.
\section{Model-size variation in single-task training and activity analysis}
\label{app:other_sizes}
\subsection{Single-task training}
\label{app:other_size_single}
\begin{figure}[htbp]
    \centering
    \includegraphics[width=1\textwidth]{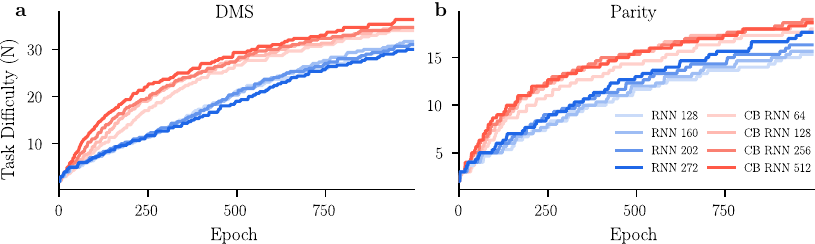}
    \vspace{-1.5em}
    \caption{\textbf{CB-RNN performance advantage is preserved across model sizes.}
    Single-task curriculum progression for DMS (a) and Parity (b) across CB-RNNs with different GC-layer sizes and approximately parameter-matched RNN-only baselines. CB-RNNs generally progressed faster and reached higher task difficulty across model scales.}
    \label{fig:more_gcs}
\end{figure}

Single-task training was repeated across CB-RNNs with GC-layer sizes of 64, 128, 256, and 512, using 256 in the main analysis, alongside parameter-matched RNN-only baselines. Across both DMS and Parity, CB-RNNs generally progressed faster through the curriculum and reached higher task difficulty than matched RNN-only models (Fig.~\ref{fig:more_gcs}). Thus, the learning advantage was preserved across model capacities and was not driven by a single choice of CB module size.

To test whether GC-layer expansion contributed to the CB-RNN advantage, we compared the AUC difference between each CB-RNN and its parameter-matched RNN-only baseline for a non-expansive CB module ($GC=64$) and the main expanded model ($GC=256$). The AUC advantage was larger for the expanded GC model, particularly on DMS, suggesting that expansion in the CB pathway contributes to the performance benefit.
\begin{figure}[htbp]
    \centering
    \includegraphics[width=1\textwidth]{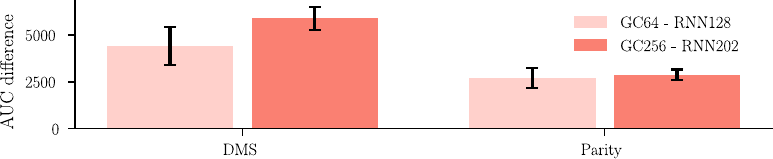}
    \vspace{-1.5em}
    \caption{\textbf{GC-layer expansion increases the CB-RNN AUC advantage.}
    Difference in curriculum AUC between CB-RNNs and their approximately parameter-matched RNN-only baselines for non-expansive ($GC=64$) and expanded ($GC=256$) CB modules. The expanded GC model showed a larger AUC advantage, most clearly on DMS. Bars show mean AUC difference; error bars show SD across random seed repeats.}
    \label{fig:auc_difs}
\end{figure}
\subsection{Representational Analyses}
\label{app:other_size_reps}
\begin{figure}[htbp]
    \centering
    \includegraphics[width=1\textwidth]{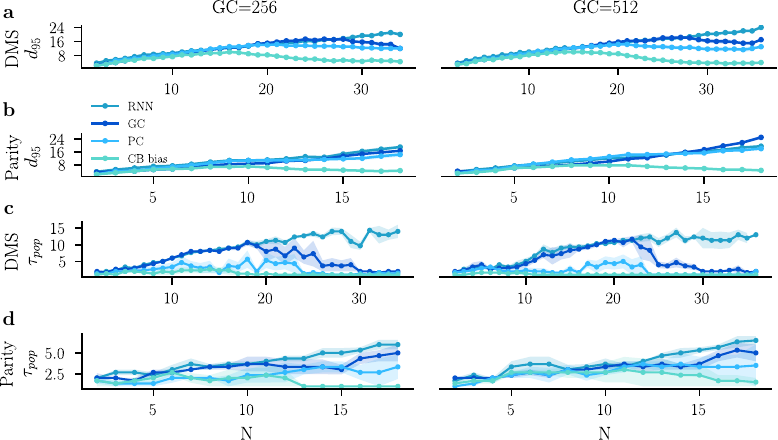}
    \vspace{-1.5em}
    \caption{\textbf{Representational metrics are broadly preserved across GC expansion sizes.}
    (a, b) Dimensionality ($d_{95}$) and (c, d) population timescale ($\tau_{pop}$) were compared between $GC=256$ (left) and $GC=512$ (right) CB-RNNs. 
    Across both tasks, the qualitative patterns were similar across GC sizes, indicating that the main representational results were not strongly dependent on the chosen GC expansion size. Shaded regions show SD across random seed repeats.}
    \label{fig:bigvssmall}
\end{figure}

To ensure that the representational results were not strongly dependent on the chosen GC expansion size, dimensionality and population timescale analyses were repeated for a larger $GC=512$ CB-RNN and compared with the $GC=256$ model used in the main analyses. The qualitative patterns were broadly preserved across GC sizes (Fig.~\ref{fig:bigvssmall}).

\section{Cerebellar input source controls}
\label{app:input_controls}
To test whether CB-RNN performance depended on the information available to the CB module, single-task DMS and Parity runs were repeated using restricted CB input configurations. In the main analyses, the CB module received the concatenated recurrent state and task input, $[h_t;x_t]$. Control models instead received either the recurrent hidden state alone ($h_t$ only; no direct task input) or the task input alone ($x_t$ only; no recurrent state input).
\begin{figure}[htbp]
    \centering
    \includegraphics[width=1\textwidth]{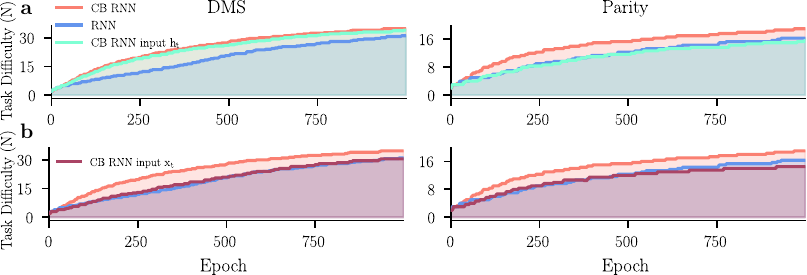}
    \vspace{-1.5em}
    \caption{\textbf{CB-RNN performance depends on the information available to the cerebellar module.}
    Single-task DMS and Parity runs were repeated with restricted CB input configurations. (a) When CB input is only hidden state activity $h_t$, DMS performance is largely preserved but Parity performance falls to match baseline. (b) When CB input is restricted to task input $x_t$ only, performance across DMS and Parity falls to match baseline. Lines reflect mean across random seeds, shading indicates AUC.}
    \label{fig:input_comparisons}
\end{figure}
The effect of removing CB access to hidden state input $h_t$ was task-dependent, with performance largely preserved in DMS, but falling to match RNN-only in Parity (Fig.~\ref{fig:input_comparisons}a). When removing task input $x_t$, performance across both tasks fell to be similar to RNN-only baseline, performing slightly worse than the baseline in Parity (Fig.~\ref{fig:input_comparisons}b). These results suggest that recurrent state information is particularly important for the CB contribution to DMS, while Parity benefits from access to both recurrent state and the current task input.
\section{Task generation details}
\label{app:task_generation}
\paragraph{Single task}
All tasks used binary input sequences $x_t \in \{0,1\}$ generated online during 
training and evaluation. For each batch, a sequence length $M$ was sampled as a 
function of the largest active curriculum level $N_{\max}$:
\[
M \in \{N_{\max}+2,\ldots,4N_{\max}+1\}.
\]
Longer curriculum levels were thus paired with 
longer input sequences, ensuring the final target always had sufficient temporal 
context. Input bits were sampled independently from a balanced Bernoulli distribution,
$x_t \sim \mathrm{Bernoulli}(0.5)$, and each batch contained $B=64$ independently 
generated sequences arranged as a tensor $X \in \mathbb{R}^{M \times B \times 1}$.

\paragraph{N-bit delayed match-to-sample.}
The label for a sequence of length $M$ was a single binary value indicating whether 
the final input bit matched the bit $N-1$ timesteps earlier:
\[
y^{\mathrm{DMS}}_N = [s_{t-N+1} = s_t],
\]
where $t = M$ denotes the final timestep.

\paragraph{N-bit parity.}
The label was the parity of the final $N$ input bits:
\[
y^{\mathrm{Parity}}_N = \left(\sum_{i=0}^{N-1} x_{M-i}\right) \bmod 2.
\]

\paragraph{Batch outputs.}
Each task-generation call returned the input tensor $X \in \mathbb{R}^{M \times B 
\times 1}$ and a list of label tensors, one per active curriculum level $N$, each 
of shape $B$. 

\begin{algorithm}[H]
\caption{CB-RNN curriculum training protocol.}
\label{alg:training}

\KwData{Task $\mathcal{T} \in \{\mathrm{DMS}, \mathrm{Parity}\}$, initial difficulty $N=2$, epoch budget $E_{\max}=1000$, advancement threshold $\tau=0.98$}

\KwInit{Initialise RNN weights, CB weights, readout heads, and optimiser}

\For{epoch $e = 1,\ldots,E_{\max}$}{

    \tcp{Training phase}
    \For{gradient step $g = 1,\ldots,100$}{
        Sample sequence length $M$ as a function of current difficulty $N$\;
        Generate batch $X \in \{0,1\}^{M \times B \times 1}$ with $B=64$\;
        Compute task labels $y_N$ for the current difficulty level\;

        \For{timestep $t = 1,\ldots,M$}{
            Compute CB bias $b_{\mathrm{cb},t} = \mathrm{CB}([h_{t-1};x_t])$\;
            Update recurrent state:
            \[
            h_t = (1-\alpha)h_{t-1} + \alpha[W_hh_{t-1}+W_xx_t+b+b_{\mathrm{cb},t}]_{\gamma}
            \]
        }

        Compute readout prediction from final hidden state $h_M$\;
        Compute cross-entropy loss $\mathcal{L}$\;
        Backpropagate through the unrolled sequence\;
        Clip gradients to $\ell_2$ norm $\leq 7.5$\;
        Update trainable parameters with SGD and Nesterov momentum\;
    }

    \tcp{Evaluation and curriculum advancement}
    Evaluate mean accuracy $\hat{a}$ on 50 newly generated batches\;

    \If{$\hat{a} \geq \tau$}{
        Save checkpoint for current $N$\;
        Increase difficulty: $N \leftarrow N+1$\;
    }

    \If{$N > 100$}{
        \textbf{break}\;
    }
}
\end{algorithm}
\paragraph{Multi-task}
For multi-task training, the same binary input batch was used for DMS and Parity. Separate labels were generated for each task at the current shared curriculum level, and separate task-specific readout heads produced the corresponding predictions. The training loss was the sum of the DMS and Parity cross-entropy losses. Curriculum advancement occurred only when both task accuracies exceeded 98\% on evaluation batches.
\paragraph{Task-switching batches.}
In task-switching experiments, only the currently active task contributed to the training loss. The active task-specific readout head and shared network parameters were updated during that phase, while both task heads were evaluated after each epoch. Switches occurred when the active task reached the next multiple of $N_{\mathrm{switch}}=5$, and the incoming task resumed at that inherited difficulty.

\section{PCA for population dimensionality analysis}
\label{app:pca_details}
Activity was extracted separately for each recorded population: the recurrent hidden state, GC layer, PC layer, and CB bias signal. For each population, activity was stored as a tensor of shape $T \times B \times D$, where $T$ is sequence length, $B$ is batch size, and $D$ is the dimensionality of the population. Activity was flattened across the temporal and batch dimensions to produce a samples-by-units matrix $X \in \mathbb{R}^{TB \times D}$. Each matrix was mean-centred across samples before PCA was applied separately for each population, checkpoint, task, and random seed. PCA was fit without specifying a fixed number of components, so all available components up to $\min(TB,D)$ were retained before dimensionality metrics were computed.

Dimensionality was summarised as $d_{95}$, the smallest number of principal components required to explain 95\% of the variance. This threshold-based measure provides an interpretable estimate of the number of dimensions required to capture the dominant population activity structure.

\section{Cerebellar ablation analyses}
\subsection{t-SNE implementation details}
\label{app:tsne_details}
To visualise the effect of CB pathway ablation on class-discriminative recurrent representations, t-SNE was applied to recurrent hidden-state activity from representative checkpoints. For DMS, the checkpoint at $N=12$ was used; for Parity, the checkpoint at $N=6$ from run 2 was used. Recurrent activity was extracted from the intact CB-RNN and from the same model after post-training silencing of the CB bias. Activity was projected into two dimensions using t-SNE with two components, perplexity 30, PCA initialisation, automatic learning-rate selection, and random seed 0.

\subsection{Quantification of class separability}
\label{app:class_separability}
To quantify the separation shown in the t-SNE visualisations, class separability was measured in the hidden state space. For each independently trained network, a linear logistic-regression decoder was trained to classify trial identity from final-timestep hidden activity using stratified five-fold cross-validation. Decoder performance was summarised as the mean ROC-AUC across folds for each run, and values are reported as mean $\pm$ SD across random seed repeats.
\begin{table}[htbp]
\centering
\caption{\textbf{Linear separability of hidden-state representations after CB ablation.}
Class separability was quantified from final-timestep hidden activity using a linear logistic-regression decoder with stratified five-fold cross-validation. Values report mean ROC-AUC $\pm$ SD across random seeds.}
\label{tab:hidden_separability_ablation}
\begin{tabular}{llcc}
\toprule
Task & $N$ & Full model & CB ablated \\
\midrule
DMS & $12$ & $0.998 \pm 0.000$ & $0.582 \pm 0.004$ \\
Parity & $6$ & $0.997 \pm 0.004$ & $0.671 \pm 0.069$ \\
\bottomrule
\end{tabular}
\end{table}
Given the reduction in ROC-AUC values following CB ablation across DMS and Parity tasks, we suggest that the CB pathway is facilitating the formation or maintenance of linearly decodable task representations in the recurrent hidden state.

\subsection{Readout retraining during ablation}
\label{app:readout_retraining}
To test whether the performance loss following CB pathway ablation could be explained solely by a mismatch between recurrent representations and the trained output readout, we performed a readout-only recovery analysis. Ablations were repeated as in Methods~\ref{methods:ablation}, but with readout training enabled. Performance was compared across three conditions: the intact CB-RNN, the CB-disabled model before readout retraining, and the CB-disabled model after readout-only retraining.

\begin{figure}[htbp]
    \centering
    \includegraphics[width=1\textwidth]{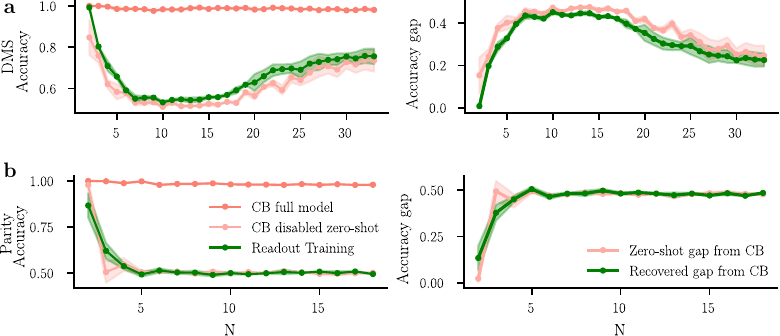}
    \vspace{-1.5em}
    \caption{\textbf{Readout-only retraining does not rescue cerebellar pathway ablation.}
    (a) In DMS, disabling the CB pathway strongly reduced accuracy, and retraining only the output readout produced only partial recovery. (b) In Parity, readout retraining produced little recovery after CB pathway ablation. In both tasks, the residual gap from the intact CB-RNN remained substantial, indicating that the CB pathway supports task-relevant recurrent representations. Lines show mean across runs; shaded regions indicate SEM.}
    \label{fig:readout_recovery}
\end{figure}

Across both DMS and parity, disabling the CB pathway caused a large drop in accuracy relative to the intact CB-RNN (Fig.~\ref{fig:readout_recovery}). Readout-only retraining failed to restore performance to the intact model level. In DMS, readout retraining produced partial recovery at larger $N$s, as expected given the findings with readout heads frozen, but a substantial gap from the full CB-RNN remained across task difficulties. In Parity, readout retraining provided little additional recovery beyond the zero-shot CB-disabled condition, with both remaining close to chance-level performance across most $N$s. Thus, the behavioural impairment caused by CB pathway ablation cannot be explained simply by a poorly matched output readout.

\section{Performance statistics}
\label{app:performance-statistics}
Descriptive statistics were calculated across repeat runs with random seeds to demonstrate performance differences.  
\begin{table}[h]
\centering
\caption{\textbf{Curriculum-learning performance for model variants under single-task and multi-task training.} Values are mean $\pm$ SD across runs. Epochs to $N_{\mathrm{global}}/2$ is measured against the global maximum $N$ across all models within each setting.}
\label{tab:performance_stats_table}
\begin{tabular}{llccc}
\toprule
Setting & Model & Max $N$ & Epochs to $N_{\mathrm{global}}/2$ & AUC \\
\midrule
\multirow{4}{*}{\shortstack[l]{Single-task\\DMS}}
  & CB-RNN               & $34.67 \pm 0.58$ & $217.00 \pm 7.00$   & $25{,}162 \pm 203$ \\
  & Interleaved Reservoir & $35.00 \pm 3.00$ & $269.00 \pm 19.52$  & $23{,}383 \pm 236$ \\
  & Full Reservoir        & $29.33 \pm 1.53$ & $288.33 \pm 7.51$   & $21{,}303 \pm 508$ \\
  & RNN-only             & $31.00 \pm 0.00$ & $443.00 \pm 39.51$  & $19{,}282 \pm 757$ \\
\midrule
\multirow{4}{*}{\shortstack[l]{Single-task\\Parity}}
  & CB-RNN               & $19.00 \pm 0.00$ & $146.33 \pm 9.71$   & $14{,}330 \pm 213$ \\
  & Interleaved Reservoir & $16.33 \pm 1.15$ & $175.33 \pm 10.02$  & $12{,}528 \pm 534$ \\
  & Full Reservoir        & $15.33 \pm 0.58$ & $170.33 \pm 9.24$   & $12{,}161 \pm 267$ \\
  & RNN-only             & $16.33 \pm 0.58$ & $300.33 \pm 37.87$  & $11{,}458 \pm 404$ \\
\midrule
\multirow{4}{*}{Multi-task}
  & CB-RNN               & $22.33 \pm 0.58$ & $157.00 \pm 7.94$   & $16{,}590 \pm 132$ \\
  & Interleaved Reservoir & $18.33 \pm 2.31$ & $221.00 \pm 18.03$  & $14{,}232 \pm 315$ \\
  & Full Reservoir        & $18.00 \pm 1.73$ & $238.00 \pm 35.68$  & $13{,}599 \pm 597$ \\
  & RNN-only             & $16.67 \pm 1.53$ & $539.00 \pm 28.05$  & $10{,}939 \pm 511$ \\
\bottomrule
\end{tabular}
\end{table}
\clearpage


\end{document}